\documentclass[%
 reprint,
superscriptaddress,
 amsmath,amssymb,
 aps,
prstab,
floatfix,
]{revtex4-2}

\usepackage{graphicx}
\usepackage{dcolumn}
\usepackage{bm}
\usepackage{amsmath}
\usepackage{siunitx}
\usepackage{hyperref}

\newcommand{\pyht}{\textsc{pyHeadTail~}}

\begin{document}

\title{Self-consistent derivation of the transverse mode coupling instability for coasting beams using the linearized Vlasov equation}


\author{N.~Biancacci}
\affiliation{%
CERN, Geneva, Switzerland
}%
\author{E.~M\'etral}
\affiliation{%
CERN, Geneva, Switzerland
}%
\author{M.~Migliorati}
\affiliation{%
University of Rome, ``La Sapienza'' and INFN - Roma 1, Rome, Italy
}%

\date{\today}

\begin{abstract}
The mode coupling instability for coasting beams has been discussed in a previous paper using macroparticle tracking simulations from the \pyht code and a simple analytical formula which was proposed as an extension of the ansatz used for the single-particle formalism. In this paper, we propose a self-consistent derivation of this formula based on the linearized Vlasov equation. The proposed mode coupling instability for coasting beams was never predicted or discussed in the past and we believe that the reason is twofold. First, to derive it analytically from the linearized Vlasov equation, one should not make the usual approximation $\sin(\phi)\simeq (e^{j\phi})/(2j)$, where $\phi$ is the transverse betatron phase, but really consider the two terms of $\sin(\phi)=(e^{j\phi}-e^{-j\phi})/(2j)$ as the second term is the one responsible for the mode coupling in coasting beams. It should be stressed here that mode coupling is found already with driving impedance only. Note that the previous approximation is also usually made for bunched beams and this case should therefore also be carefully reviewed in the future. Second, by including the detuning impedance, the coupling is much stronger and this is what we found also in \pyht simulations. 
\end{abstract}

\maketitle


\section{Introduction}
\label{sec:intro}

\begin{figure}[htb!]
	\centering
	\includegraphics[width=1\columnwidth]{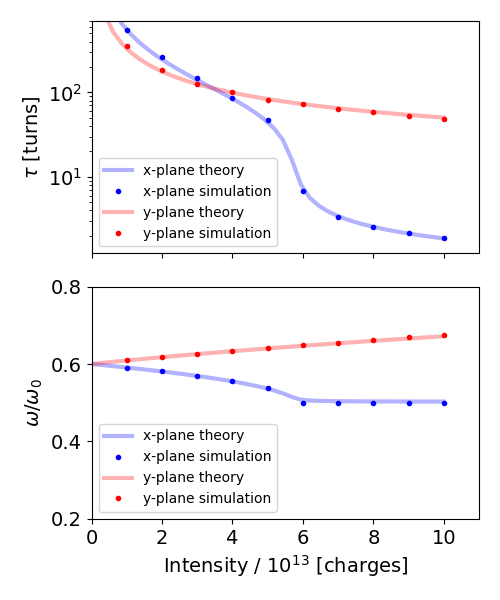}
	\caption{Same figure as Fig. 6 of~\cite{PhysRevAccelBeams.23.124402}, using a logarithmic scale for the vertical axis of the instability rise times. \pyht simulations (with dots) compared to theory accounting for coupling between two waves (with full lines) for the horizontal (blue) and the vertical (red) planes. The rise time, at the top, and the normalized frequency shift, at the bottom, are shown.}
\label{fig:ZoomedPicture}
\end{figure}

The mode coupling instability for coasting beams was first discussed in a recent paper~\cite{PhysRevAccelBeams.23.124402}, with a four-step approach. First, we extended the \pyht simulation code to simulate transverse (and longitudinal) coherent instabilities for coasting beams. Second, to gain confidence in what had been done, we first benchmarked the new simulations with the classical transverse coasting beam approach from Laclare~\cite{bib:laclare_coast} and an excellent agreement was reached, as can be seen in Figs. 3 of~\cite{PhysRevAccelBeams.23.124402}, for both the real and imaginary parts of the complex tune shift. Third, we introduced the detuning impedance in \pyht simulations with coasting beams and found that the results did not agree anymore with the classical approach. Equation (23) of~\cite{PhysRevAccelBeams.23.124402} was then
proposed as an extension of Chao’s ansatz (see Eq.~(5.71) of page 243 of Chao’s textbook~\cite{bib:chao_book}). This equation was found to be in excellent agreement with the new \pyht simulations, as can be observed in Fig.~6 of~\cite{PhysRevAccelBeams.23.124402}, nicely reproducing, in particular, the plane exchange of the most critical instability vs.~intensity (which was one of the important findings of the study). The picture is reproduced here in Fig.~\ref{fig:ZoomedPicture}, using a logarithmic scale for the vertical axis of the instability rise times, to better reveal the excellent agreement between the new theory and the \pyht simulations over all the intensity range. Fourth, as this new Eq. (23) of~\cite{PhysRevAccelBeams.23.124402} was not derived (yet) self-consistently from the Vlasov equation, we mentioned in~\cite{PhysRevAccelBeams.23.124402} that “This result is presently being investigated following also the Vlasov’s formalism and will allow future studies on the effect of a finite momentum spread and chromaticity”. The study of the effect of a finite momentum spread and chromaticity is still work in progress, but we provide here the full derivation of Eq. (23) (or Eq. (24)) of~\cite{PhysRevAccelBeams.23.124402} using the linearized Vlasov approach.

\section{Mode coupling instability with driving impedance only}

In this section, we will follow the approach (and notations) of Laclare~\cite{bib:laclare_coast} and extend his analysis without performing the approximation $\sin(\phi)\simeq (e^{j\phi})/(2j)$, with $j$ the imaginary unit, but we will consider the two terms of $\sin(\phi)=(e^{j\phi}-e^{-j\phi})/(2j)$. Furthermore, we will study the mode $p$ of the slow wave (in our definition, this corresponds to the wave oscillating as $e^{+jw_{pc}t}$, with $w_{pc}$ the coherent angular frequency for mode $p$ we are looking for and $t$ the time) and the mode $q$ of the fast wave (in our definition, this corresponds to the wave oscillating as $e^{-jw_{qc}t}$, with $w_{qc}$ the coherent angular frequency for mode $q$ we are looking for).
The distribution function will thus be written
\begin{equation}
\label{eq:1}
    \Psi = \Psi_0+\Psi_p 
\end{equation}
for the slow wave and
\begin{equation}
\label{eq:2}
    \Psi = \Psi_0+\Psi_q
\end{equation}
for the fast wave, where  
\begin{equation}
\label{eq:3}
    \Psi_0=g_0(\dot{\tau})f_0(\hat{x}) 
\end{equation}
is the stationary distribution, while
\begin{equation}
\label{eq:4}
    \Psi_p=g_1(\dot{\tau})f_1(\phi,\hat{x})e^{j(w_{pc}t+p\omega_0\tau)}
\end{equation}
is the perturbation for the slow wave and 
\begin{equation}
\label{eq:5}
    \Psi_q=g_1(\dot{\tau})f_1(\phi,\hat{x})e^{j(-w_{qc}t+q\omega_0\tau)}
\end{equation}
is the perturbation of the fast wave. Here, $\tau$ is the time delay between the reference particle and the test particle (at the same point in the machine), $\dot{\tau}$ is the derivative of $\tau$ versus time, $\hat{x}$ is the transverse betatron amplitude and $\omega_0$ is the angular revolution frequency. In the absence of chromaticity and spreads, the transverse betatron phase is written (with $Q$ the transverse tune)
\begin{equation}
\label{eq:6}
    \phi=Q\theta,
\end{equation}
where $\theta$ is the azimuthal angle given by
\begin{equation}
\label{eq:7}
    \theta=\omega_0(t-\tau).
\end{equation}
The evolution of $\hat{x}$ can be expressed as a function of the transverse (driving) impedance as defined in~\cite{bib:laclare_coast} (but multiplied by $\beta$) and, for the slow wave, it is given by
\begin{equation}
\label{eq:8}
    \dot{\hat{x}}=\dfrac{\sin(\phi)}{\dot{\phi}}\dfrac{e I}{m_0\gamma\beta c}jZ_pe^{j[(p\omega_0+w_{pc})t-p\theta]}J,
\end{equation}
while it is given by 
\begin{equation}
\label{eq:9}
    \dot{\hat{x}}=\dfrac{\sin(\phi)}{\dot{\phi}}\dfrac{e I}{m_0\gamma\beta c}jZ_qe^{j[(q\omega_0-w_{qc})t-q\theta]}J
\end{equation}
for the fast wave. Here, $e$ is the elementary charge, $I$ is the beam current, $m_0$ is the rest mass of the particles, $\gamma$ and $\beta$ are the relativistic mass and velocity factors, $c$ is the speed of light, $J$ is the dipole moment and $Z_p$ and $Z_q$ are the transverse (driving) impedances evaluated at the frequency $(p+Q)\omega_0$ for the slow wave and $(q-Q)\omega_0$ for the fast wave. With our formalism, the waves corresponding to  $p=-7$ and $q=6$ (with $Q=6.4$) correspond to the case discussed in~\cite{PhysRevAccelBeams.23.124402}, which was also compared to \pyht simulations. 

The Vlasov equation is written
\begin{equation}
\label{eq:10}
    0=\dfrac{d\Psi}{dt}=\dfrac{\partial{\Psi}}{\partial{t}}+\dfrac{\partial{\Psi}}{\partial{\phi}}\dot{\phi}+\dfrac{\partial{\Psi}}{\partial{\hat{x}}}\dot{\hat{x}}+\dfrac{\partial{\Psi}}{\partial{\tau}}\dot{\tau},
\end{equation}
and it will be solved below for both the slow and fast waves, considering the following distribution functions for the perturbations
\begin{equation}
\label{eq:11}
    g_1(\dot{\tau})=g_0(\dot{\tau})=\dfrac{\omega_0}{2 \pi}\delta(\dot{\tau}),
\end{equation}
where $\delta$ is the Dirac function,
\begin{equation}
\label{eq:12}
    f_1(\phi,\hat{x})=-J\dfrac{2\pi}{\omega_0}\dfrac{\partial{f_0(\hat{x})}}{\partial{\hat{x}}}e^{-j \phi}
\end{equation}
for the slow wave and
\begin{equation}
\label{eq:13}
    f_1(\phi,\hat{x})=-J\dfrac{2\pi}{\omega_0}\dfrac{\partial{f_0(\hat{x})}}{\partial{\hat{x}}}e^{j \phi}
\end{equation}
for the fast wave. 

Starting with the mode $p$ of the slow wave, the (linearized) Vlasov equation becomes
\begin{multline}
\label{eq:14}
    0=j w_{pc} g_1(\dot{\tau})f_1(\phi,\hat{x})e^{j(p\omega_0\tau+w_{pc}t)} \\
    +g_1(\dot{\tau})\dfrac{\partial{f_1(\phi,\hat{x})}}{\partial{\phi}}\dot{\phi}e^{j(p\omega_0\tau+w_{pc}t)} \\
    +g_0(\dot{\tau})\dfrac{\partial{f_0(\hat{x})}}{\partial{\hat{x}}}\dfrac{\sin(\phi)}{\dot{\phi}}\dfrac{e I}{m_0\gamma\beta c}jZ_pe^{j[(p\omega_0+w_{pc})t-p\theta]}J.
\end{multline}
Replacing $g_1$ and $f_1$ by their definitions from Eqs. (11) and (12), yields 
\begin{multline}
\label{eq:15}
    0=-J\dfrac{\partial{f_0(\hat{x})}}{\partial{\hat{x}}}e^{-j \phi}jw_{pc}e^{j(p\omega_0\tau+w_{pc}t)} \\
    +jJ\dfrac{\partial{f_0(\hat{x})}}{\partial{\hat{x}}}e^{-j \phi}\dot{\phi}e^{j(p\omega_0\tau+w_{pc}t)} \\
    +\dfrac{\partial{f_0(\hat{x})}}{\partial{\hat{x}}}\dfrac{\sin(\phi)}{\dot{\phi}}\dfrac{e I \omega_0}{2\pi m_0\gamma\beta c}jZ_pe^{j[(p\omega_0+w_{pc})t-p\theta]}J.
\end{multline}
Removing the common terms $J\dfrac{\partial{f_0(\hat{x})}}{\partial{\hat{x}}}$ yields
\begin{multline}
\label{eq:16}
    0=-e^{-j \phi}jw_{pc}e^{j(p\omega_0\tau+w_{pc}t)} \\
    +je^{-j \phi}\dot{\phi}e^{j(p\omega_0\tau+w_{pc}t)} \\
    +\dfrac{\sin(\phi)}{\dot{\phi}}\dfrac{e I \omega_0}{2\pi m_0\gamma\beta c}jZ_pe^{j[(p\omega_0+w_{pc})t-p\theta]}.
\end{multline}
Replacing now $\dot{\phi}$ by $Q\omega_0$ and $\sin(\phi)$ by $(e^{j\phi}-e^{-j\phi})/(2j)$ gives 
\begin{multline}
\label{eq:17}
    0=-e^{-j \phi}jw_{pc}e^{j(p\omega_0\tau+w_{pc}t)} \\
    +je^{-j \phi}e^{j(p\omega_0\tau+w_{pc}t)}Q\omega_0 \\
    +\dfrac{e I}{2\pi Q m_0\gamma\beta c}jZ_pe^{j[(p\omega_0+w_{pc})t-p\theta]}\dfrac{(e^{j\phi}-e^{-j\phi})}{2j}.
\end{multline}
Removing the $j$ and combining the terms, one obtains
\begin{multline}
\label{eq:18}
    e^{-j \phi}e^{j(p\omega_0\tau+w_{pc}t)}[w_{pc}-Q\omega_0-\dfrac{e I}{4\pi Q m_0\gamma\beta c}jZ_p] \\
    =-\dfrac{e I}{4\pi Q m_0\gamma\beta c}jZ_pe^{j(p\omega_0\tau+w_{pc}t)}e^{j \phi},
\end{multline}
which can also be written
\begin{multline}
\label{eq:19}
    e^{j(p\omega_0\tau+w_{pc}t)}[w_{pc}-Q\omega_0-\dfrac{e I}{4\pi Q m_0\gamma\beta c}jZ_p] \\
    =-\dfrac{e I}{4\pi Q m_0\gamma\beta c}jZ_pe^{j(p\omega_0\tau+w_{pc}t)}e^{2j \phi}.
\end{multline}
As $w_{pc}=Q\omega_0+\Delta w_{pc}$, one can write
\begin{multline}
\label{eq:20}
    e^{j(p\omega_0\tau+w_{pc}t)}e^{2j \phi}=e^{j(p+Q)\omega_0 t}e^{j\Delta w_{pc}t}e^{-jp\theta}e^{2jQ\theta} \\
    =e^{j(q-Q)\omega_0 t}e^{j\Delta w_{pc}t}e^{-jp\theta}e^{j(p-q+2Q)\omega_0 t}e^{2jQ\theta},
\end{multline}
and therefore the final equation for the slow wave is given by
\begin{multline}
\label{eq:21}
    e^{j(p+Q)\omega_0 t}e^{j\Delta w_{pc}t}e^{-jp\theta}[w_{pc}-Q\omega_0-\dfrac{e I}{4\pi Q m_0\gamma\beta c}jZ_p] \\
    =-\dfrac{e I}{4\pi Q m_0\gamma\beta c}jZ_pe^{j(q-Q)\omega_0 t}e^{j\Delta w_{pc}t}e^{-jp\theta}  \\
    e^{j(p-q+2Q)\omega_0 t}e^{2jQ\theta},
\end{multline}
which can be simplified as
\begin{multline}
\label{eq:22}
    e^{j(p+Q)\omega_0 t}[w_{pc}-Q\omega_0-\dfrac{e I}{4\pi Q m_0\gamma\beta c}jZ_p] \\
    =-\dfrac{e I}{4\pi Q m_0\gamma\beta c}jZ_pe^{j(q-Q)\omega_0 t}  \\
    e^{j(p-q+2Q)\omega_0 t}e^{2jQ\theta}.
\end{multline}
Redoing now exactly the same derivation for the fast wave, one obtains the following final equation
\begin{multline}
\label{eq:23}
    e^{j(q-Q)\omega_0 t}[w_{qc}-Q\omega_0-\dfrac{e I}{4\pi Q m_0\gamma\beta c}jZ_q] \\
    =-\dfrac{e I}{4\pi Q m_0\gamma\beta c}jZ_qe^{j(p+Q)\omega_0 t} \\
    e^{-j(p-q+2Q)\omega_0 t}e^{-2jQ\theta}.
\end{multline}
Combining the 2 coupled equations from Eq.(22) and Eq.~(23) (by multiplying both sides), one obtains the equation describing the coupling between a slow wave and a fast wave
\begin{multline}
\label{eq:24}
    [w_{pc}-Q\omega_0-\dfrac{e I}{4\pi Q m_0\gamma\beta c}jZ_p][w_{qc}-Q\omega_0-\dfrac{e I}{4\pi Q m_0\gamma\beta c}jZ_q] \\
    =\dfrac{e I}{4\pi Q m_0\gamma\beta c}jZ_p\dfrac{e I}{4\pi Q m_0\gamma\beta c}jZ_q.
\end{multline}
Note that another way to derive Eq. (24) is to start from the slow-wave Eq. (19) and remove on both sides the same exponential term: one ends up with only the $e^{2j \phi}$ term on the right-hand side. Doing the same thing with the fast-wave Eq. (23), one ends up with only the $e^{-2j \phi}$ term on the right-hand side. Multiplying then both equations, Eq. (24) is recovered, revealing clearly that the coupling terms between the two equations are $e^{\pm 2j \phi}$.
Defining for the slow wave $\Omega=p\omega_0+w_{pc}$, in the absence of mode coupling, $\Omega=(p+Q)\omega_0+\Delta w_{pc}$ with \begin{equation}
\label{eq:25}
    \Delta w_{pc}=\dfrac{e I}{4\pi Q m_0\gamma\beta c}jZ_p.
\end{equation}
Similarly for the fast wave, one can define $\Omega=q\omega_0-w_{qc}$ and in the absence of mode coupling, $\Omega=(q-Q)\omega_0-\Delta w_{qc}$ with \begin{equation}
\label{eq:26}
    \Delta w_{qc}=\dfrac{e I}{4\pi Q m_0\gamma\beta c}jZ_q.
\end{equation}
Using these definitions, Eq. (24) can be written as
\begin{multline}
\label{eq:27}
    [\Omega-(p+Q)\omega_0-\Delta w_{pc}] \\
    [-\Omega+(q-Q)\omega_0-\Delta w_{qc}]=\Delta w_{pc}\Delta w_{qc},
\end{multline}
which is the final equation we were looking for. Indeed, Eq. (27) is exactly the same as the equation obtained from Eq. (24) of~\cite{PhysRevAccelBeams.23.124402}, when the determinant of the matrix is equal to zero and when the following fully justified approximations are made there: $\Omega\simeq\ n_1 \omega_0 - \omega_{\beta}$ in $m_{11}$ and $\Omega\simeq\ n_2 \omega_0 + \omega_{\beta}$ in $m_{22}$. The term $2 \omega_{\beta}$ can be then simplified in the four terms of the matrix and Eq. (27) is recovered. Solving this equation in the particular case discussed in~\cite{PhysRevAccelBeams.23.124402} yields exactly the same results.

\section{Mode coupling instability with both driving and detuning impedances}

In this section, we include also the detuning impedance and we show that in the absence of spreads, the contribution from the detuning impedance just adds to the one from the driving impedance and therefore Eq.~\eqref{eq:27} can still be used with the total impedance being defined as the sum of the driving impedance and the detuning impedance (with the latter being evaluated at zero frequency).

In presence of detuning impedance alone, the evolution of $\hat{x}$ is modified as

\begin{multline}
\label{eq:28}
 \dot{\hat{x}}=j  \dfrac{\sin(\phi)}{\dot{\phi}} \dfrac{e}{m_0\gamma}  \dfrac{1}{2\pi R}\\ \left (x \int_{-\infty}^{+\infty} Z^{det}(\omega) S_\parallel (\omega,\theta) e^{j \omega t} d \omega \right ),
\end{multline}
where $S_\parallel (\omega,\theta)$  is the total longitudinal signal induced by the beam at azimuth position $\theta$. If the beam is longitudinally stable, the signal is $S_\parallel (\omega,\theta) =  I \delta(\omega)$ and 
\begin{equation}
    \label{eq:29}
    \dot{\hat{x}}= \dfrac{\sin(\phi)}{\dot{\phi}}\dfrac{e I}{m_0\gamma \beta c} \left [j \dfrac{\omega_0}{2\pi} x Z^{det} (0)\right ].    
\end{equation}
The equation of motion of a particle under the effect of a detuning impedance alone is given by~\cite{bib:laclare_coast}
\begin{equation}
    \label{eq:30}
    \Ddot{x} + \dot{\phi}^2 x = -\dfrac{e I}{m_0\gamma \beta c}\left [j \dfrac{\omega_0}{2\pi} x Z^{det} (0)\right ].
\end{equation}
The betatron frequency will change according to \mbox{$\dot{\phi}^\prime = \dot{\phi} +\Delta \dot{\phi}$}, with $\Delta \dot{\phi}$ given, for small deviations from the unperturbed betatron frequency, by 
\begin{equation}
    \label{eq:31}
    \Delta \dot{\phi} = j \dfrac{e I}{4\pi Q m_0\gamma \beta c}Z^{det}(0).
\end{equation}

In the presence of both driving and detuning impedance, one needs to account in Eq~\eqref{eq:30} the additional contribution of the driving impedance. For the slow wave, Eq.~\eqref{eq:30} would be modified as
\begin{multline}
    \label{eq:32}
    \Ddot{x} + \dot{\phi}^2 x =- j\dfrac{e I}{m_0\gamma  \beta c}\\
    \left [\dfrac{\omega_0}{2\pi} x Z^{det} (0) + Z_p e^{j[(p\omega_0+w_{pc})t-p\theta]}J \right ].
\end{multline}

For consistency, the $x$ time dependence must follow the one of the driving impedance, i.e. 
\begin{equation}
\label{eq:33}
    x(t)= A e^{j(p\omega_0+w_{pc})t},
\end{equation} 
where $A$ is a complex amplitude coefficient that can be found by computing the total transverse signal at the pickup $S_\perp(t,\theta)$ in two ways. The first one, also performed in~\cite{bib:laclare_coast}, is given by
\begin{align}
\label{eq:34}
    S_\perp(t,\theta) &= \int_v \Psi(\tau, \dot{\tau}, \phi, \hat{x},t)\cdot  x \cdot s_\parallel(t,\theta) \, \mathrm{d}v \\
    &= \int_v \Psi_p(\tau, \dot{\tau}, \phi, \hat{x},t) \cdot \hat{x} \cos{\phi} \cdot s_\parallel(t,\theta) \, \mathrm{d}v \\
    &= \dfrac{2\pi I}{\omega_0}  J e^{j[(p \omega_0 + w_{pc} )t} e^{-p\theta]},
\end{align}
with $\mathrm{d}v=\mathrm{d}\tau \mathrm{d}\dot{\tau}\hat{x}\mathrm{d}\hat{x}\mathrm{d}\phi$ infinitesimal phase space volume and $s_\parallel(t,\theta)$ the longitudinal signal of a particle at the pickup
\begin{equation}
\label{eq:35}
    s_\parallel(t,\theta) = \dfrac{e \omega_0}{2\pi} \sum_p e^{j p[\omega_0 (t-\tau)-\theta ]}.  
\end{equation}
The second one computes the signal using Eq.~\eqref{eq:33}:
\begin{align}
\label{eq:36}
    S_\perp(t,\theta) &= \int_v \Psi(\tau, \dot{\tau}, \phi, \hat{x},t)\cdot  x \cdot s_\parallel(t,\theta) \, \mathrm{d}v  \\
    &= \int_v \Psi_0(\dot{\tau}, \hat{x}) A e^{j(p\omega_0+w_{pc})t}  s_\parallel(t,\theta)\,\mathrm{d}v \\
    &= A I e^{j(p \omega_0+w_{pc})t}.
\end{align}
Comparing Eqs.(36) and (40) we can conclude that 
\begin{equation}
\label{eq:37}
    A = \dfrac{2\pi}{\omega_0} J  e^{-j p\theta }. 
\end{equation}
Combining Eqs.~\eqref{eq:33} and~\eqref{eq:37} we have
\begin{equation}
    \label{eq:38}
    x(t) =  \dfrac{2\pi}{\omega_0} J e^{j[(p\omega_0+w_{pc})t-p\theta]}.
\end{equation}

In the presence of both driving and detuning impedances, Eq.~\eqref{eq:29} therefore becomes
\begin{multline}
    \label{eq:39}
    \dot{\hat{x}}=\dfrac{\sin(\phi)}{\dot{\phi}}\dfrac{e I}{m_0\gamma  \beta c} j \\ \left  [  Z^{det} (0) + Z_p  \right] e^{j[(p\omega_0+w_{pc})t-p\theta]} J,
\end{multline}
showing that the detuning impedance, sampled at DC, adds up to the driving one. Equation~\eqref{eq:39} is indeed the same equation as Eq.~\eqref{eq:8} with the total impedance replacing the driving impedance.

\section{Conclusions}
\label{sec:conclusions}
The purpose of this paper was to derive self consistently, using the linearized Vlasov equation and following Laclare's formalism~\cite{bib:laclare_coast}, Eqs. (23) and (24) which were proposed in~\cite{PhysRevAccelBeams.23.124402} as an extension of Chao’s ansatz (see Eq. (5.71) of page 243 of Chao’s textbook~\cite{bib:chao_book}). The exact equations were recovered, as can be seen, for example, from Eq. (27), confirming the existence of this new instability mechanism: the transverse mode coupling instability for coasting beams. 

This equation was already found to be in excellent agreement with the \pyht simulations discussed in~\cite{PhysRevAccelBeams.23.124402} (see Fig. 6). This picture was reproduced in Fig.~\ref{fig:ZoomedPicture}, using a logarithmic scale for the vertical axis of the instability rise times, to reveal this agreement between the new theory and the \pyht simulations over all the intensity range.

\bibliography{biblio}{}
\bibliographystyle{IEEEtran}
\end{document}